\documentclass[conference]{IEEEtran}
\IEEEoverridecommandlockouts
\usepackage{graphicx}
\usepackage{soul}
\usepackage{epsf}
\usepackage{amsmath, amsfonts}
\usepackage{amssymb}
\usepackage{latexsym}
\usepackage{subcaption}
\usepackage{multirow}
\usepackage{multicol}
\usepackage[table]{xcolor}
\usepackage{textcomp}
\usepackage{optidef}
\usepackage[ruled,vlined]{algorithm2e}
\usepackage{algorithmic}

\DeclareMathOperator*{\argmin}{argmin}
\def\BibTeX{{\rm B\kern-.05em{\sc i\kern-.025em b}\kern-.08em T\kern-.1667em\lower.7ex\hbox{E}\kern-.125emX}}
\usepackage{balance}

\usepackage{caption}
\usepackage{subcaption}

\title{Geometric Machine Learning for Channel Covariance Estimation in Vehicular Networks}
\begin{document}
\author{
\IEEEauthorblockN{Imtiaz Nasim and Ahmed S. Ibrahim}
\IEEEauthorblockA{Department of Electrical and Computer Engineering, Florida International University, Miami, FL, USA\\
\{inasi001, aibrahim\}@fiu.edu}}

\maketitle
\begin{abstract}
Learning the covariance matrices of spatially-correlated wireless channels, in millimeter-wave (mmWave) vehicular communication, can be utilized in designing environment-aware beamforming codebooks. Such channel covariance matrices can be represented on non-Euclidean Riemannian manifolds, thanks to their symmetric positive definite (SPD) characteristics. Consequently in this paper, we propose a Riemannian-Geometric machine learning (G-ML) approach for estimating the channel covariance matrices based on unsupervised K-Means model. %, so that a matched beamforming codebook can be constructed accordingly. 
%More specifically, we applied a geometric K-means using the Log-Euclidean metric (LEM) to cluster the symmetric positive definite (SPD) covariance matrices of the fading channels that lie on the Riemannian manifold.
The proposed K-means algorithm utilizes Log-Euclidean metric (LEM) as the distance measure among channel covariance matrices over the Riemannian manifolds. We show that our proposed K-Means G-ML model can achieve up to $80\%$ less error compared to Euclidean-based K-Means algorithm, which applies clustering on the channel vectors themselves. %Our results show that utilizing the underlying geometric nature of the spatially correlated channels can provide significant performance gain compared to those based on the direct channel vectors for learning the covariance functions. Moreover, the proposed unsupervised approach makes the solution suitable for a high dynamic vehicular environment without requiring any prior training.
\end{abstract}

%\vspace{0.2 in}


\IEEEpeerreviewmaketitle

%%%%%%%%%%%%%%%%%%%%%%%%%%%%%%%%%%%%%%%%%%%%%%%%%%%%%%%%%%%%%%%%%%%%%%%%%%%%%%%%%%%%%%%%%%%%
% Introduction
%%%%%%%%%%%%%%%%%%%%%%%%%%%%%%%%%%%%%%%%%%%%%%%%%%%%%%%%%%%%%%%%%%%%%%%%%%%%%%%%%%%%%%%%%%%%
\section{Introduction}\label{sec_intro}
The increasing demand for high data rates to support emerging applications (e.g., vehicular communication) has induced numerous research challenges in high mobility networks \cite{wu2016}. Millimeter-Wave (mmWave) is a promising candidate for these applications to support high rates~\cite{va_survey_16}, thanks to its large bandwidth availability \cite{rappaport13}. %However, before enabling mmWave for dynamic environments, such as vehicular communications, proper understanding of such environments is necessary for precise channel characterization between the transmitter and its user equipment (UE) \cite{colo_19}. Due to rapid change in positions of vehicular UEs, the environment changes very fast that can lead to significant degradation in the system performance. Such varying nature of the vehicular environment not only makes the beam selection challenging \cite{imtiaz_20}, but also makes it difficult to determine what specific codebook should be used at the transmitter to match its local environment. In particular, 
In a dynamic vehicular environment, the transmission from a multiple-antenna base station (or generally a relay) to a vehicular user equipment (UE) depends on the spatially correlated channels \cite{adhikary_13} which are determined by their relative positions. Since each relay’s location is different, each relay will experience different channel characteristics towards its UEs. As such, having a fixed pre-defined beamforming codebook for all relays will be inefficient. Instead, there is a need for each relay to learn its wireless environment
characteristics (i.e., channel covariance matrices), then construct a matched beamforming codebook according to~\cite{love_06,choi_16,raghavan_07}. Learning the spatially correlated covariance matrices can dramatically reduce the network overhead needed to design an environment-aware beamforming codebook~\cite{raghavan_07}.  %Consequently, we need to develop efficient techniques to acquire the mmWave channels in order to obtain the channel knowledge.  

There are many studies in the literature that estimated the covariance matrices due to their advantages towards obtaining the channel knowledge \cite{hu_19,neumann_18}. The work in \cite{hu_19} investigated the covariance estimation problem for hybrid mmWave systems by forming a low-rank matrix structure to minimize the complexity. The authors in \cite{neumann_18} proposed covariance estimation by adopting a maximum-likelihood technique based on systematic allocations of pilot sequences. The work in \cite{park_18} performed a compressive sensing based covariance estimation for sparse mmWave channels. Nevertheless, there is a need for leveraging machine learning (ML) algorithms that can dynamically adapt to the changes in a mmWave vehicular environment \cite{alkhateeb_20} for predicting the channel covariance matrices. 

A recent study in \cite{alkhateeb_asilomar18} used a supervised ML approach that required prior training with label information to correctly estimate the covariance matrices. However, such supervised solutions may not be feasible for high dynamic vehicular scenarios, especially when the network is equipped with mobile base stations and labelled samples for prior training are not available. Therefore, we need a learning strategy without prior label information that can precisely estimate the covariance matrices to meet the high-dynamic vehicular demand. To do this, we turn our attention to unsupervised ML that requires \emph{no labelled training} \cite{unsupervise_online} to learn the channel covariance matrices.
%and allows the model to learn on its own by identifying the unknown structure within the given data \cite{unsupervise_online}. 

The success of any unsupervised ML significantly depends on identifying the underlying structure within the dataset under consideration \cite{unsupervise_online}. In that matter, we note that the mmWave multiple-antenna channels have spatial correlation characteristics and their covariance matrices are symmetric positive semi-definite ones in nature. Applying positive shifts to these semi-definite matrices transform them to symmetric positive definite (SPD) ones which can be modeled as points over the conic (i.e., non-euclidean) Riemannian manifolds~\cite{ibrahim_globecom20,2015_Sra_Cone_SPD}. Making use of these manifolds can actually help in accurate estimation of the channel covariance matrices, which we will show later in this paper.

Riemannian geometry \cite{2019_Intro_RiemGeometry} has been recently considered in designing beamforming vectors~\cite{Fan19,chen17}. Additionally, non-euclidean methods have been utilized in codebook design such as \emph{non-conic} Riemannian manifolds (i.e, unitary, fixed rank) as in~\cite{2019_Schober_BF_Riem,2016_Letaief_Riem_mmWave_Precoding}. While these studies present novel non-euclidean geometric perspectives of beamforming and codebook designs, they have not been utilized in channel covariance estimation. %Also, Riemannian metrics, such as Log-Euclidean metric (LEM), has been previously used for task classification \cite{arsingy_06}. Non-euclidean metrics like LEM consider the geodesic curves while calculating the distance between two points over the  Riemannian manifold. 
Consequently, our goal is to make use of the underlying Riemannian-geometric structures of the \emph{spatially-correlated} multiple-antenna channels for channel covariance estimation in vehicular networks.

%%%%%%%%%%%%%%%%%%%%%%%%%%%
Geometric machine learning (G-ML) algorithms can characterize the data given in non-euclidean domain which is not possible with conventional machine learning approaches that deal with euclidean-structured data samples~\cite{monti_16,bronstein_17}.
In this paper, we propose an unsupervised approach, namely Geometric K-Means ML, to cluster the SPD variants of the channel covariance matrices with distinguishable spatial-correlation characteristics. Then Riemannian metrics, such as Log-Euclidean metric (LEM)~\cite{arsingy_06}, is applied to calculate the geodesic distance among these SPD covariance matrices over the given Riemannian manifold. We show that our proposed non-euclidean based G-ML approach can provide higher accuracy in clustering the channel covariance matrices compared to the euclidean-based solutions. The contributions of this paper can be highlighted as follows:
\begin{itemize}
    \item For estimating the covariance matrices in a mmWave vehicular network, we model the problem as an unsupervised clustering problem that requires no prior label information while training to learn the covariance matrices as opposed to the existing supervised solutions.
    \item We propose  a Geometric K-Means ML, that makes use of the underlying Riemannian-geometric characteristics of the spatially correlated multiple-antenna channels in estimating the channel covariance matrices.
\end{itemize}

The rest of the paper is organized as follows. Section~\ref{sec_sys_model} introduces the system model along with some preliminary concepts. Section~\ref{sec_scheme} introduces the proposed solution and Section~\ref{sec_results} shows the simulation results. Finally, Section~\ref{sec_conclusion} concludes the paper.
%%%%%%%%%%%%%%%%%%%%%%%%%%%%%%%%%%%%%%%%%%%%%%%%%
%%%%%%%%%%%%%%%%%%%%%%%%%%%%%%%%%%%%%%%%%%%%%%%%%%%
\section{System Model}\label{sec_sys_model}
In this section, we first provide a brief preliminary on Riemannian geometry and then describe the system model for covariance estimation in a mmWave vehicular network.
%%%%%%%%%%%%%%%%
\begin{figure}[b]
\centering
\includegraphics[width = .6\linewidth, height = 1.65 in]{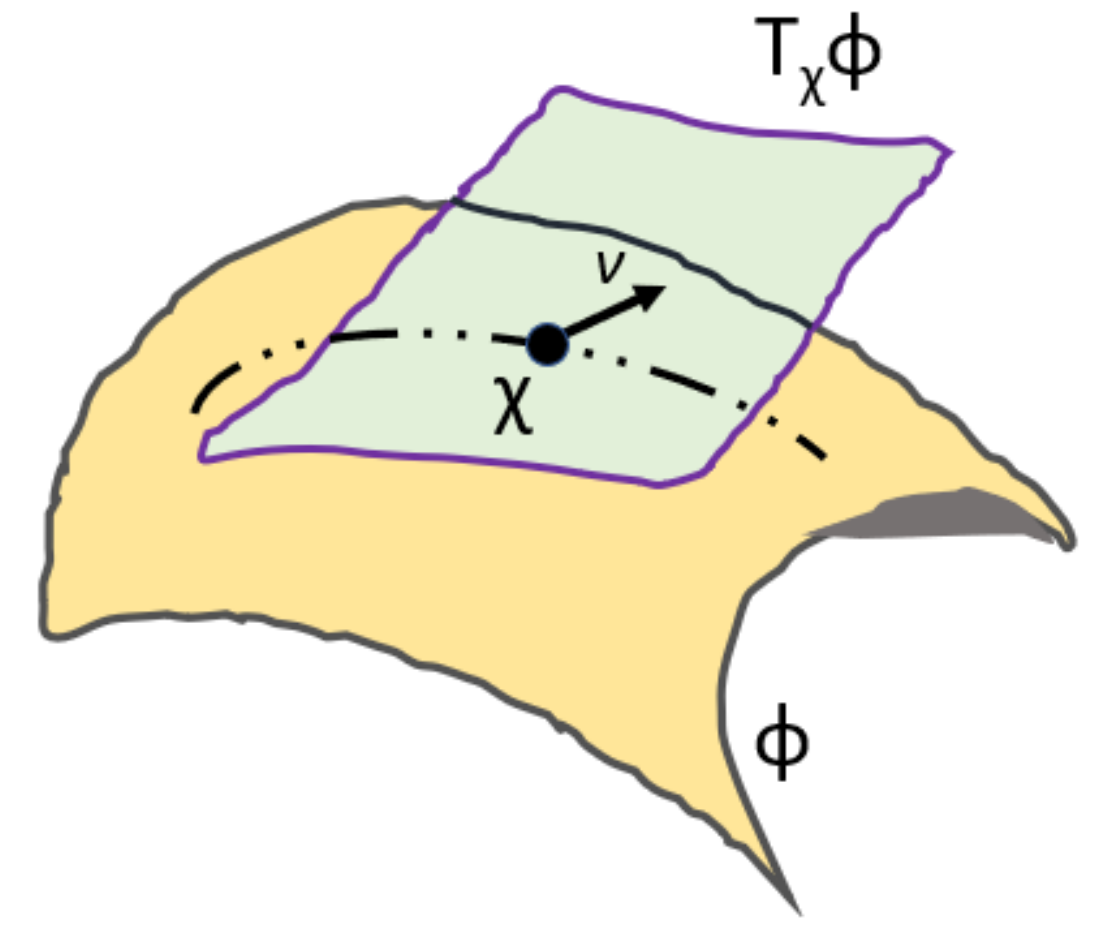}
\caption{\small Non-Euclidean manifold $\mathbf{\Phi}$ and its $2$-dimensional tangent plane T$_\chi\,\mathbf{\Phi}$ at point $\chi$, which is a collection of all tangent vectors $\nu$.}
\label{fig_manifold}
\end{figure}
%%%%%%%%%%%%%%%%%%%%%%%%%%%%
\subsection{Riemannian Geometry}
A non-euclidean manifold  $\mathbf{\Phi}$, is illustrated in Fig. \ref{fig_manifold}.
The tangent space T$_\chi\, \mathbf{\Phi}$ at a given point $\chi$ is composed of all the vectors $\nu$ which are tangent to the manifold $\mathbf{\Phi}$ through the point $\chi$. We define an exponential map as the function mapping of a point $\chi$ on the tangent space T$_\chi\, \mathbf{\Phi}$ of manifold $\mathbf{\Phi}$, exp$_\chi$ : T$_\chi\, \mathbf{\Phi}$ → $\mathbf{\Phi}$.
In most cases, manifolds are classified as a special category of topological manifolds, namely Riemannian manifolds~\cite{2019_Intro_RiemGeometry,2016_Letaief_Riem_mmWave_Precoding}. SPD matrices lie on the interior of such Riemannian manifolds which have a cone-shaped structure~\cite{2015_Sra_Cone_SPD}. Riemannian metrics, such as LEM~\cite{arsingy_06}, are inner products on the tangent space T$_\chi\, \mathbf{\Phi}$ and measure the length of geodesics i.e., shortest length of curves over the manifold among the given SPD matrices.

%%%%%%%%%%%%%%%%%%%%%%%%%%%%%%%%%%
\subsection{Vehicular Communications}
\begin{figure}[h]
\centering
\includegraphics[width = .9\linewidth, height=1.65 in]{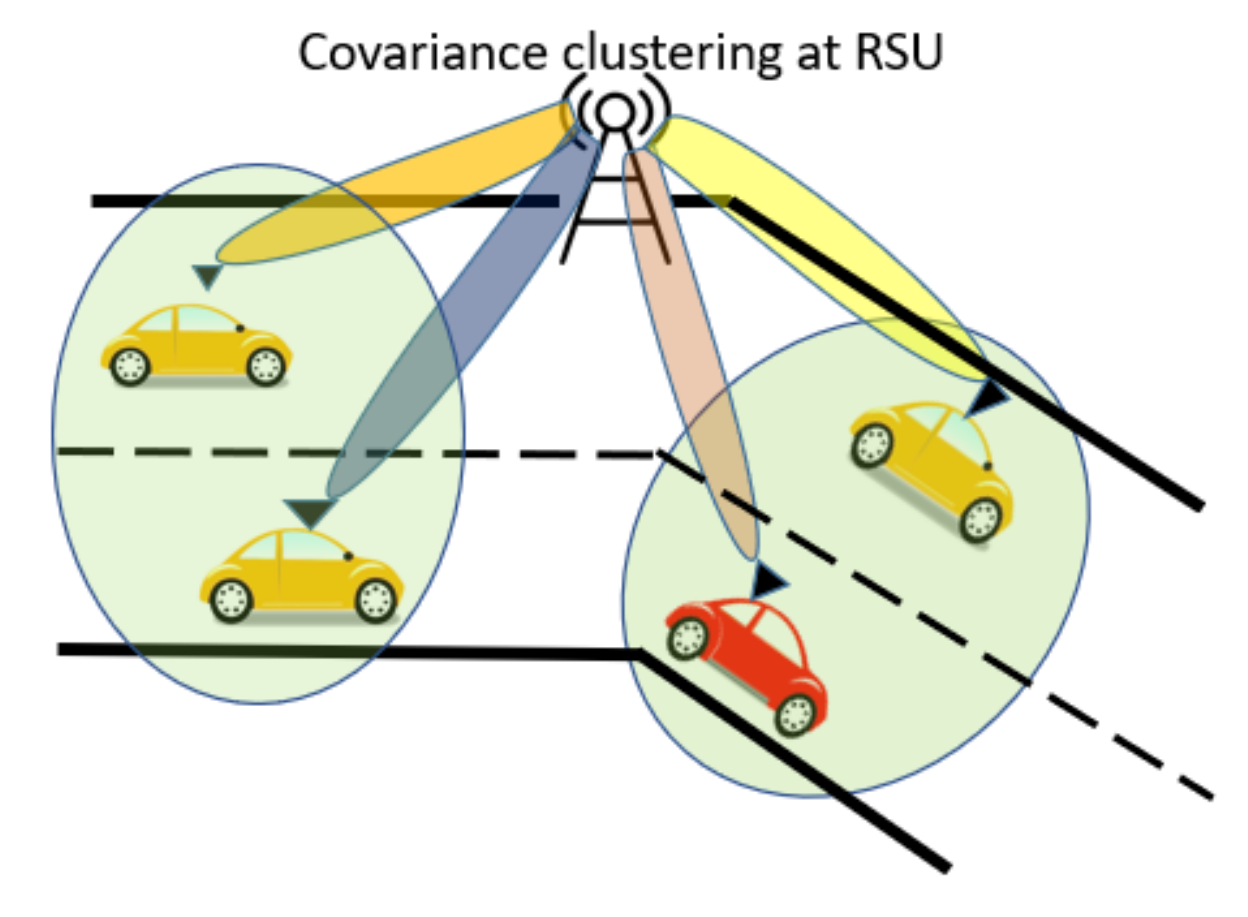}
\caption{A vehicular scenario with users located at different locations.}
\label{fig_rsu}
\end{figure}

We show a typical vehicular scenario in Fig. \ref{fig_rsu} where the users are located at different locations of a street. The different locations resemble different vehicular environments that have unique channel characteristics and varies fast as the vehicle moves. For clustering the covariance matrices in such a scenario, we consider a base station or road side unit (RSU) with $m$ antennas and a single antenna at each UE. 
%%%%%%%%%%%%%%%%%%%%%%%%%%%%%%%%%%%%%%
%%%%%%%%%%%%%%%%%%%%%%%%%%%%%%%%%%%%%%%%%%%%
Let $\mathbf{h_i}$ $\in$  $\mathbb{C}^{m\times 1}$
represents an $m \times 1$ correlated wireless channel vector between the base station and one of its UE at the $i-$th location spot. A multiple-input single-output (MISO) channel between a given base station and its UE is modeled as a correlated Rayleigh fading channel vector \cite{ibrahim_globecom20}.
Calculating the spatial correlation of these multi-antenna vehicular channels results in symmetric positive semi-definite covariance matrices. Applying positive shift to these matrices transforms them into SPD ones according to the following 
\begin{align}\label{spd}
\mathbf{Q}_i = \lambda \mathbf{I} + \mathbf{h}_i\mathbf{h}^H_i \, , 
\end{align}
where $\mathbf{I}$ is a $m \times m$ identity matrix, $\lambda$ is an arbitrary small scalar, and $\textit{H}$ denotes matrix hermitian. We consider $\lambda = 0.5$ in this paper. These sample covariance matrices (or simply SPD matrices) lie on the conic Riemannian manifold over the locations $i=1,2,...,N$. Consequently, $\mathbf{\Phi} =  \{ \mathbf{Q}_{1},\mathbf{Q}_{2},...,\mathbf{Q}_{N}\}$ represents the set of observations for all the given SPD matrices,
%\begin{align}\label{spd_all}
%\end{align}
where each observation within $\mathbf{\Phi}$ is an $m\times m$ matrix.
%Therefore, the SPD covariance matrices in (\ref{spd_all}) are the ones that need to be clustered.

%These covariance matrices lie on a Riemannian conic manifold and can be clustered according to their spatial correlation characteristics. In particular, we resort to a G-ML approach, which applies the K-Means unsupervised learning to cluster the SPD variant of the user's channel over a geometric Riemannian manifold. 
%%%%%%%%%%%%%%%%%%%%%%%%%%%%%%%%%%%%%%%%%%%
%%%%%%%%%%%%%%%%%%%
\section{Learning the Channel Covariance Matrices}\label{sec_scheme}
%%%%%%%%%%%%%%%%%%%
\subsection{Problem Description}
In a high-dynamic vehicular scenario, the covariance matrices of the spatially correlated channels are not known in advance as they keep changing based on the nature of the environment. Hence,
designing fixed beamforming codebook in advance is not an efficient solution for a dynamic vehicular case. Consequently, the codebook design needs to be learned based on the user's channels within each relay’s cluster, and hence the covariance matrices.
We aim to partition the collected samples of the shifted covariance matrices into %$k=\{1,2,...,K\}$ 
$K\leq N$ disjoint groups $R_1,R_2,..., R_K$, based on their distinguishable spatial correlation characteristics. For doing so, we need to find the set of the estimated means (i.e., centroids) of each group given as
$\mathbf{\hat{G}} =  \{ \mathbf{\hat{G}}_1, \mathbf{\hat{G}}_2,. . ., \mathbf{\hat{G}}_K \}$. The set of indices for the location points that belong to any group $R_k$ is denoted as $l_k = \{1,2,...,n\}$, where $k=1,2,...,K$.
Also, we denote the channel vectors within group $R_k$ as
$\mathbf{h}_{i,k} =   \mathbf{h}_i, (\forall{i} \in l_k)$. 
For clustering the SPD covariance matrices, we first propose our non-euclidean based scheme that clusters the given samples over Riemannian manifolds. Then, to compare the performance of our proposed solution with the benchmark approaches, we clustered the covariance matrices using euclidean-based clustering technique.
%%%%%%%%%%%%%%%%%%%%%%%%%%%%%%%%%%%%%%%%%%%%%%%%
\subsection{Non-Euclidean Based Geometric Solution}\label{non_euclidean solution}
We apply an unsupervised G-ML approach, namely Geometric K-Means ML, that considers the spatially correlated samples over the Riemannian space. Since, the set of observations can be modeled on the Riemannian manifold, learning these covariance matrices can be also conducted using Riemannian metrics. We consider LEM, which is a valid metric to calculate the geodesic distance between points~\cite{ibrahim_globecom20,arsingy_06} over the Riemannian manifolds. For the given samples $\mathbf{Q}_i \in \mathbf{\Phi}$, the mean (i.e., centroid) of the channel covariance matrices over the Riemannian manifold is defined as the minimizer of sum of squared distances \cite{shinohara_10}
\begin{align}\label{Riem_centroid}
  \mathbf{\hat{G} }=  \argmin_{\mathbf{\mathbf{G} } \in \mathbf{\Phi}} \, \frac{1}{N}\sum_{i=1}^{N} {\rho}^2 (\mathbf{Q}_i, \mathbf{G})  \, ,
\end{align}
where $\rho$ is the geodesic distance calculated using LEM. The minimizer is obtained with a gradient descent procedure. From an initial value $t=0$, the optimal centroid for group $R_k$ is reached by repeatedly applying the following update formula according to \cite{pennec_04}
\begin{align}\label{cent_repeat}
  \mathbf{\hat{G}}_{k_{t+1}}=  \text{exp}_{\mathbf{\hat{G}}_{k_t}} \,\Big(\frac{1}{n} \sum_{i\in l_k} \text{log}_{\mathbf{\hat{G}}_{k_t}} (\mathbf{Q}_i) \Big) \, \, , {k = 1,2,...K} 
\end{align}
where $n$ is the number of samples within group $R_k$.
%%LEM is applied according to \cite{ibrahim_globecom20, arsingy_06} to cluster the observations in (\ref{spd_all}) with distinguishable spatial correlation characteristics.
%%%%%%%%%%%%%%%%%%%%%
\begin{figure}[t]
\centering
\includegraphics[width = .5\linewidth, height = 1.65 in]{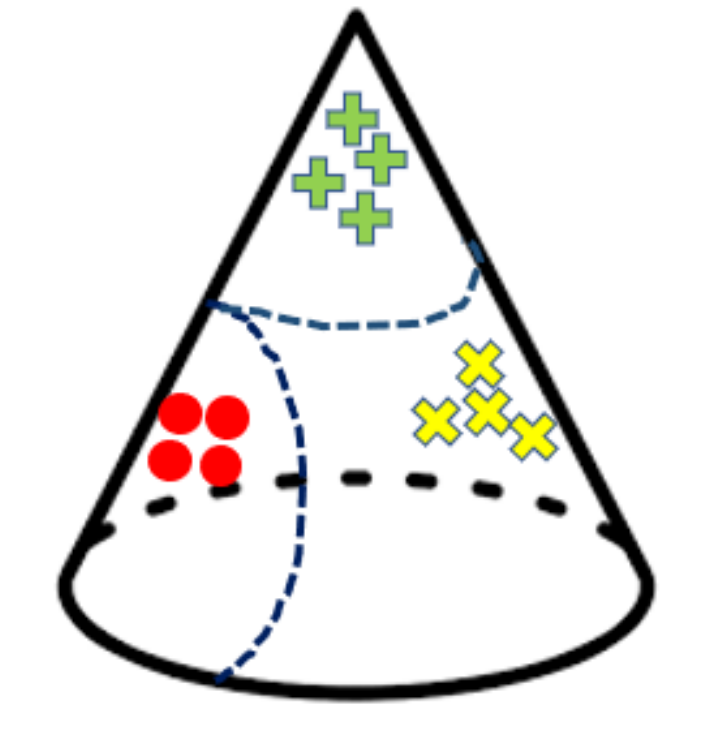}
\caption{Riemannian classification for distinguishable SPD covariance matrices.}
\label{fig_cluster}
\end{figure}
%%%%%%%%%%%%%%%%%

Finally, we subtract the induced positive shift from the estimated centroid $\mathbf{\hat{G}}_{k_{t+1}}$ to achieve the desired original estimated centroid of the covariance matrices for cluster $R_k$ according to given formula
\begin{align}\label{cov_centroid}
\mathbf{\hat{G}}_{k_c} =  \mathbf{\hat{G}}_{k_{t+1}} - \lambda \mathbf{I}. 
\end{align}
 The channel covariance matrices having the minimum LEM distances (i.e., in similar environment) from $\mathbf{\hat{G}}_{k_c}$ fall in the same group, while the samples collected from two different vehicular environments should lie in disjoint groups. Fig. \ref{fig_cluster} shows an illustration of the proposed covariance clustering scheme where the SPD covariance matrices are clustered into different groups over a conic Riemannian manifold. We use different shapes for the points to represent the distinguishable spatial correlation characteristics of the given SPD matrices in Fig. \ref{fig_cluster}.
%%%%%%%%%%%%%%

%Note that the euclidean-based K-Means algorithm will attempt to make $K$-partitions of the given dataset according to
%%%%%%%%%%%%
%\begin{mini}
%{}{\sum_{k=1}^{K} \sum_{i\in k} || \mathbf{Q}_i - %\mathbf{\hat{G} }_k ||^2}
%{\label{k_means_conv}}{} \, ,
%\end{mini}
%where $\mathbf{\hat{G} }_k$ is the estimated mean for cluster $k$. However, in a non-eucldiean manifold, we cannot apply (\ref{k_means_conv}) directly since the samples are not vectors \cite{verma_19}. Instead, we consider LEM to calculate the Riemannian-geometric distance $\rho$ among the given SPD covariance matrices and apply the following \cite{verma_19}
%%%%%%%%%%%%%
%\begin{mini}
%{\rho}{\sum_{k=1}^{K} \frac{1}{2|k|} \sum_{i,i^{'}\in k} %\rho^2 (\mathbf{Q}_i, \mathbf{Q}_ {{i^{'} } })} 
%{\label{means_geom}}{} \, .
%\end{mini}
%%%%%%%%%%%%%%%%%%

%%%%%%%%%%%%%%%%%%%%%%%%%%%%%%%%%%%%%%%%%%
\begin{figure*}[!t]
\begin{center}
\includegraphics[width=0.9\textwidth]{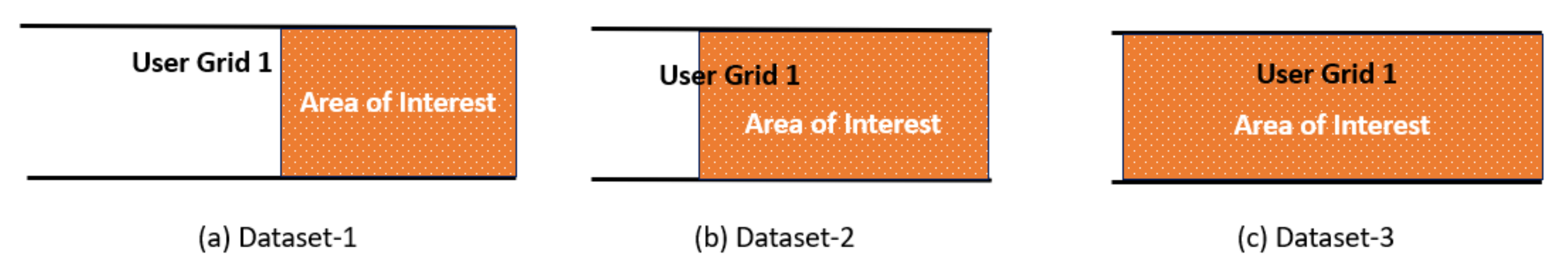}
\vskip -1ex
\caption{Datasets representing different portions of a street. \label{fig_data}}
\end{center}
\end{figure*}
%%%%%%%%%%%%%%%%%%%%%%%%%%%%
\subsection{Euclidean Based Solution}\label{euclidean soltuion}
An euclidean-based K-Means scheme takes the channel vectors $\mathbf{h}_i$ as the inputs and applies the euclidean distance to calculate the estimated centroid. Unlike our proposed solution that clusters the channel covariance matrices on Riemannian manifold, the euclidean-based K-Means attempt to make $K$-partitions of the channel vectors by applying the following minimization formula \cite{verma_19}
\begin{mini}
{\mathbf{\hat{h}}_k }{\sum_{k=1}^{K} \sum_{i\in l_k} || \mathbf{h}_i - \mathbf{\hat{h} }_k ||^2}
{\label{k_means_conv}}{} \, ,
\end{mini}
%%%%%%%%%%%%%%%%%
where $\mathbf{\hat{h}}_k$ is the centroid obtained by the euclidean-based scheme for group $R_k$. The solution then calculates the covariance matrices $\mathbf{Q}_{i,k}$ of those classified channels for each group $R_k$ as
%%%%%%%%%%%%%%%%
\begin{align}\label{eucli}
\mathbf{Q}_{i,k}= \mathbf{h}_{i,k}\mathbf{h}^H_{i,k}, \,  (\forall{i} \in l_k) \, .
\end{align}
%%%%%%%%%%%%%%

To compare the performance of our proposed G-ML solution with these benchmark euclidean-based schemes~\cite{alkhateeb_asilomar18, gao_19}, we also applied euclidean-based clustering of the channel vectors and computed their covariance matrices by applying (\ref{eucli}). Finally, the covariance matrices are predicted from these matrices of the classified samples and the estimated mean (i.e., centroid) for each group.
%%%%%%%%%%%%%%%%%%%%%%%%%%%%%%%%%%%%%%%%%%%%%%%%%%%%%%%%
%%%%%%%%%%%%%%%%%%%%%%%%%%%%%%%%%%%%%%%%%%%%%%%%%%%%%%%%
\section{Simulation Results}\label{sec_results}
In this section, we describe the different datasets that we considered along with the simulation results for learning the covariance matrices in a vehicular environment.
%%%%%%%%%%%%%%%%%%%%%%%%%%%%%%%%%%%
\subsection{Simulation Setup}\label{setup}
%%%%%%%%%%%%%%%%%%%%%%%%%%%
%{\begin{figure}
%    \centering
%    \subfigure[]{\includegraphics[width=0.29\textwidth]{data1_globecom.PNG}} 
%    \subfigure[]{\includegraphics[width=0.25\textwidth]{data2_globecom.PNG}} 
%    \subfigure[]{\includegraphics[width=0.25\textwidth]{data3_globecom.PNG}}
%    \caption{(a) Dataset-1 (b) Dataset-2 (c) Dataset-3}
%    \label{fig_data}
%\end{figure}
 
Our work assumes a vehicular network operating at 60 GHz with 0.5 GHz bandwidth and the base station is equipped with $4\times 4$ phased arrays. We adopt three different vehicular datasets from the ``$\text{O}1-60$ ray tracing scenario" \cite{alkhateeb_dataset} with their channel models. We provide brief description of these datasets before explaining the simulation results.

%%%%%%%%%%%%
%\subsubsection{Dataset-1}\label{data1}
The area of interest under Dataset-1 considers a total of 5460 user location points from user grid-1 of the ``$\text{O}1-60$ ray tracing scenario"~\cite{alkhateeb_dataset}. The area of interest for Dataset-1 is shown in Fig. \ref{fig_data} (a). 
%%%%%%%%%%%
%\subsubsection{Dataset-2}\label{data2}
For Dataset-2, our work assumes a total of 10860 location points from the considered user grid-1. The area of interest for Dataset-2 is illustrated in Fig. \ref{fig_data} (b).
%%%%%%%%%
%\subsubsection{Dataset-3} \label{data3}
The considered area under Dataset-3 is the largest one which consists of a total of 16800 samples from user grid-1 as shown in \ref{fig_data} (c). The distance between any two location points are same for all the considered scenarios. The objective for considering different datasets is to observe the efficiency of the proposed scheme for different vehicular environments.
%%%%%%%%%%%%%%%%%%%%
\subsection{Performance Analysis}\label{sec_performance}
We use the normalized mean square error (NMSE) as the considered loss function to evaluate the performance of the proposed solution for predicting the mmWave channel covariance matrices according to the following

\begin{align}\label{nmse}
\text{NMSE}_k =  \frac{||\mathbf{\hat{G}}_{k_c} - \mathbf{G}_{k_z} ||_F^2}{||\mathbf{G}_{k_z} ||_F^2} \, ,(\text{where}\, k=1,2,...,K)\,  
\end{align}
    where $\mathbf{G}_{k_z}=\mathbb{E}[\mathbf{h}_{i,k}\mathbf{h}^H_{i,k}]$ is the expected value of the covariance matrices over the location $i$ for group $R_k$, and $|| . ||_F$ denotes the matrix Frobenius norm operator. Consequently, the average NMSE over all the given clusters is denoted as
\begin{align}\label{nmse_avg}
\text{NMSE} =  \frac{1}{K} \sum_{k=1}^{K} \text{NMSE}_k \, . 
\end{align}

The proposed LEM-based Geometric K-Means ML was applied over the Riemannian manifold using the “geomstats” python package available in \cite{geomstats}. On the other hand, the euclidean-based conventional K-Means model can be applied using any of the available resources like \cite{k_means_euclidean}. We simulated the performance of the proposed non-euclidean based Geometric K-Means ML scheme and compared that with the euclidean-based solutions  for all the three considered datasets. 
%%%%%%%%%%%%%%%%%%%%%%%%%%%%%%%%%%%%%%
\begin{figure}[t]
\centering
\includegraphics[width = .8\linewidth]{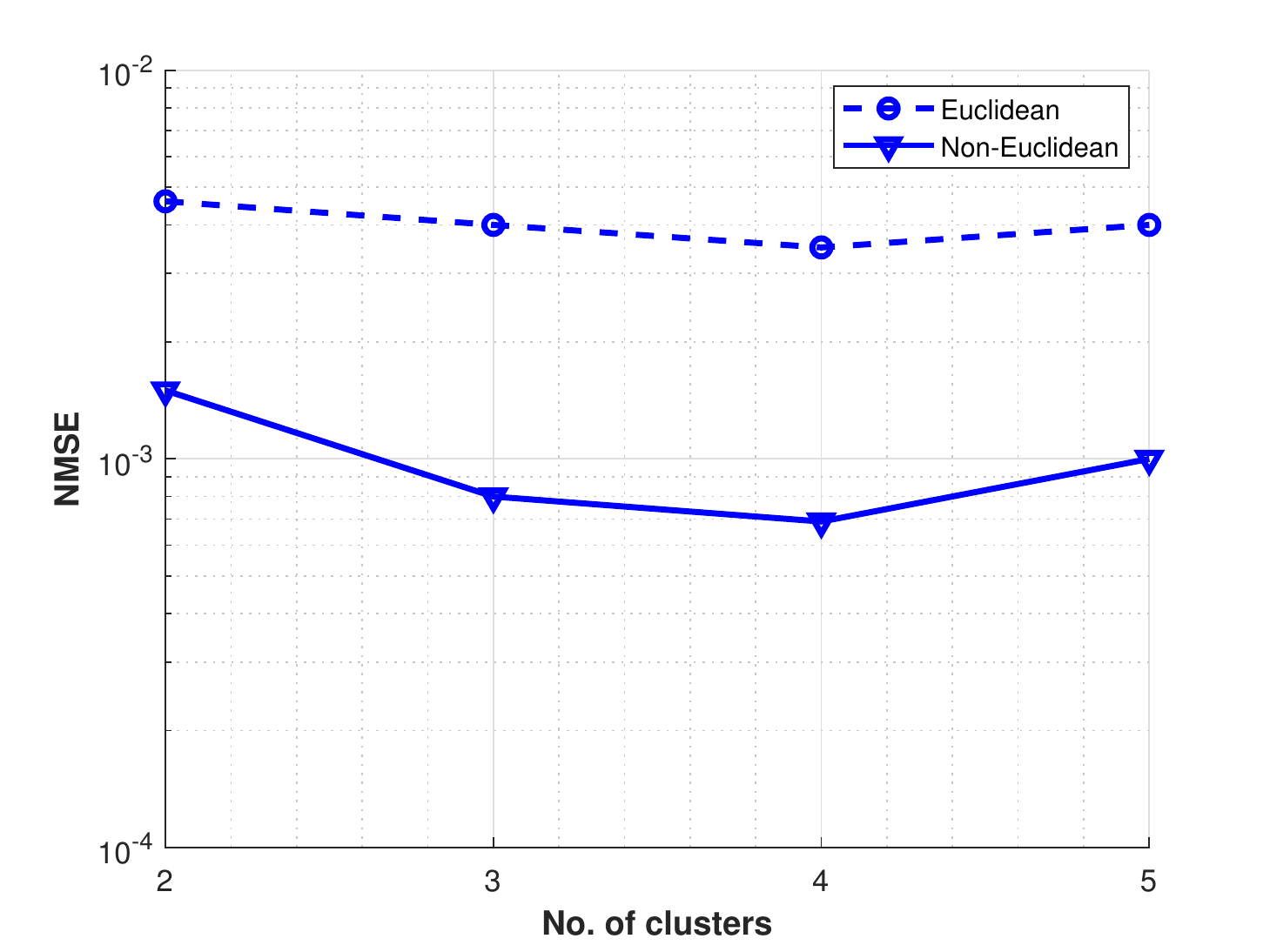}
\caption{NMSE comparison for the non-euclidean and euclidean based solution for Dataset-2.}
\label{fig_nmse_comp}
\end{figure}
%%%%%%%%%%%%%%%%%%%%%%%%%%%%%%%%%%%%%%%%%%%%%%%%%
%%%%%%%%%%%%%%%%%%%%%

Fig. \ref{fig_nmse_comp} shows the average NMSE comparison of our proposed non-euclidean based solution with the euclidean-based scheme considering Dataset-2 only. The proposed solution outperforms the euclidean scheme significantly for any given number of clusters. For example, while the euclidean-based approach provides an average NMSE value around $3.5\times 10^{-3}$ for having 4 clusters, the corresponding non-euclidean scheme achieves an NMSE around $7\times 10^{-4}$, which is 80$\%$ less than the euclidean-based solution. In other words, the proposed Geometric K-Means ML solution can provide up to 80$\%$ higher accuracy compared to the euclidean-based one. Also, the lowest NMSE value is achieved at 4 clusters for both the solutions. %Recall that our Dataset-2 considers an user array including the intersection  area as shown previously in Fig. \ref{fig_data} (b). The least NMSE at 4 clusters suggests that having 4 disjoint groups of covariance matrices should suffice the considered area of interest.
%%%%%%%%%%%%%%%%%%%%%%%%%%%%%%%%%%%

\begin{figure}[t]
\centering
\includegraphics[width = .8\linewidth]{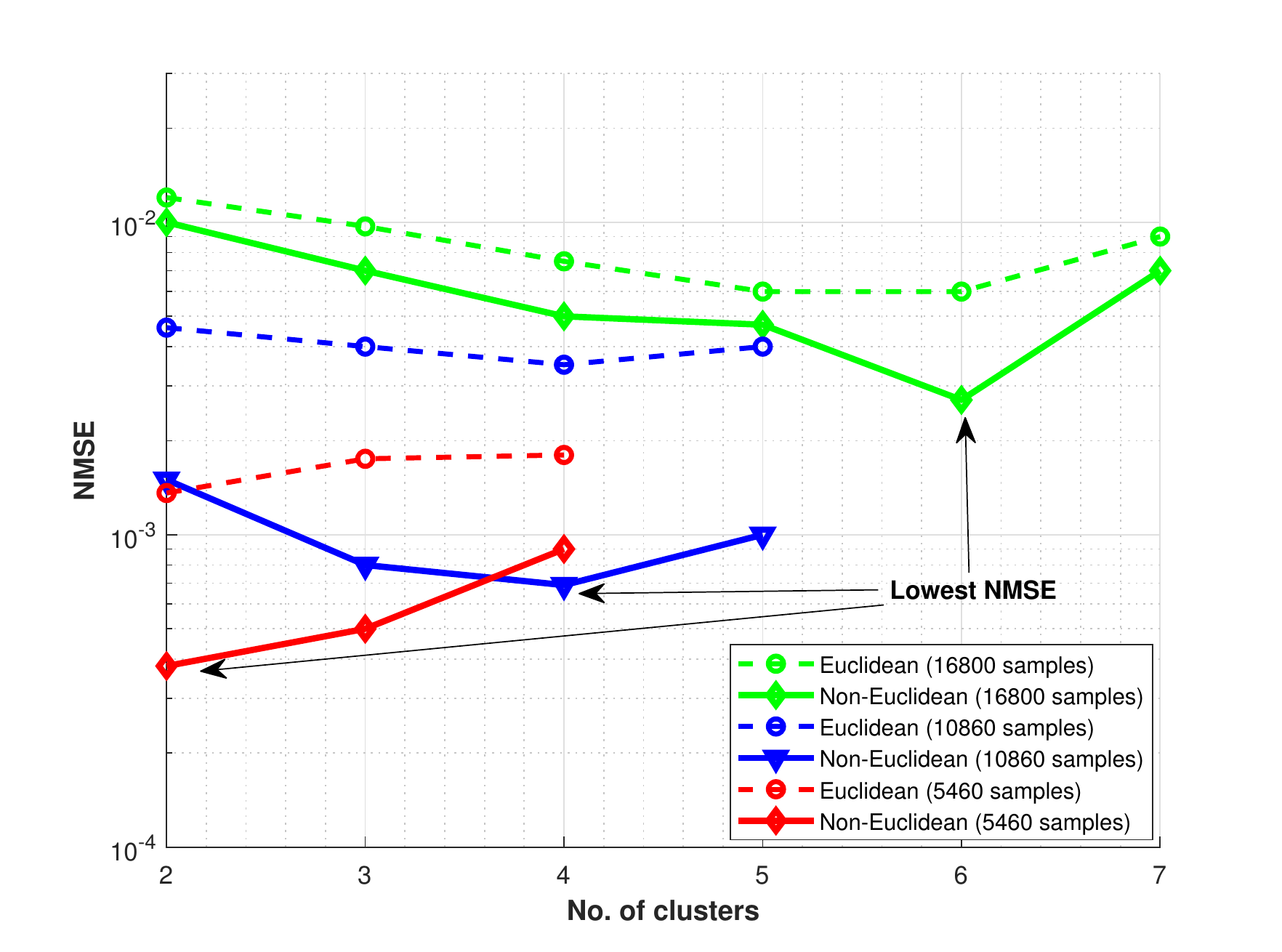}
\caption{NMSE comparison for having different data size.}
\label{fig_nmse_comp_all}
\end{figure}

Fig. \ref{fig_nmse_comp_all} shows the average NMSE performance for all the considered scenarios with Dataset-1, 2 and 3. The NMSE value is less while having a total of 5460 location points for Dataset-1, but it increases with the size of observations for Dataset-2 and 3. 
%To give an example, the proposed strategy yields an NMSE value of $0.0004$ while having 2 clusters for 5460 samples, but the NMSE for the same number of clusters with 16800 samples is $0.011$. 
This suggests a valid justification of our implementation as adding more location points should increase the NMSE. However, the proposed non-euclidean approach always performs better compared to its corresponding euclidean-based scheme for any given data size. For instance, the NMSE of the proposed solution considering 5460 samples at $4$ clusters is 50$\%$ less than that of its corresponding euclidean solution. These results indicate that clustering the channel covariance matrices directly on Riemannian manifolds based on LEM among covariance matrices is more accurate compared to clustering the channel vectors based on their euclidean distances. 
%clustering the channel covariance matrices based on their non-euclidean Riemannian-geometric distances is more accurate compared to clustering the channels based on the euclidean distances among the channels. 
Note that both of our simulated cases apply unsupervised learning and require no prior labelled training in clustering the channel covariance matrices.
%%%%%%%%%%%%%%%%%%%%%%%%%%%%%%%%%%%%%%

\begin{figure}[t]
\centering
\includegraphics[width = .8\linewidth]{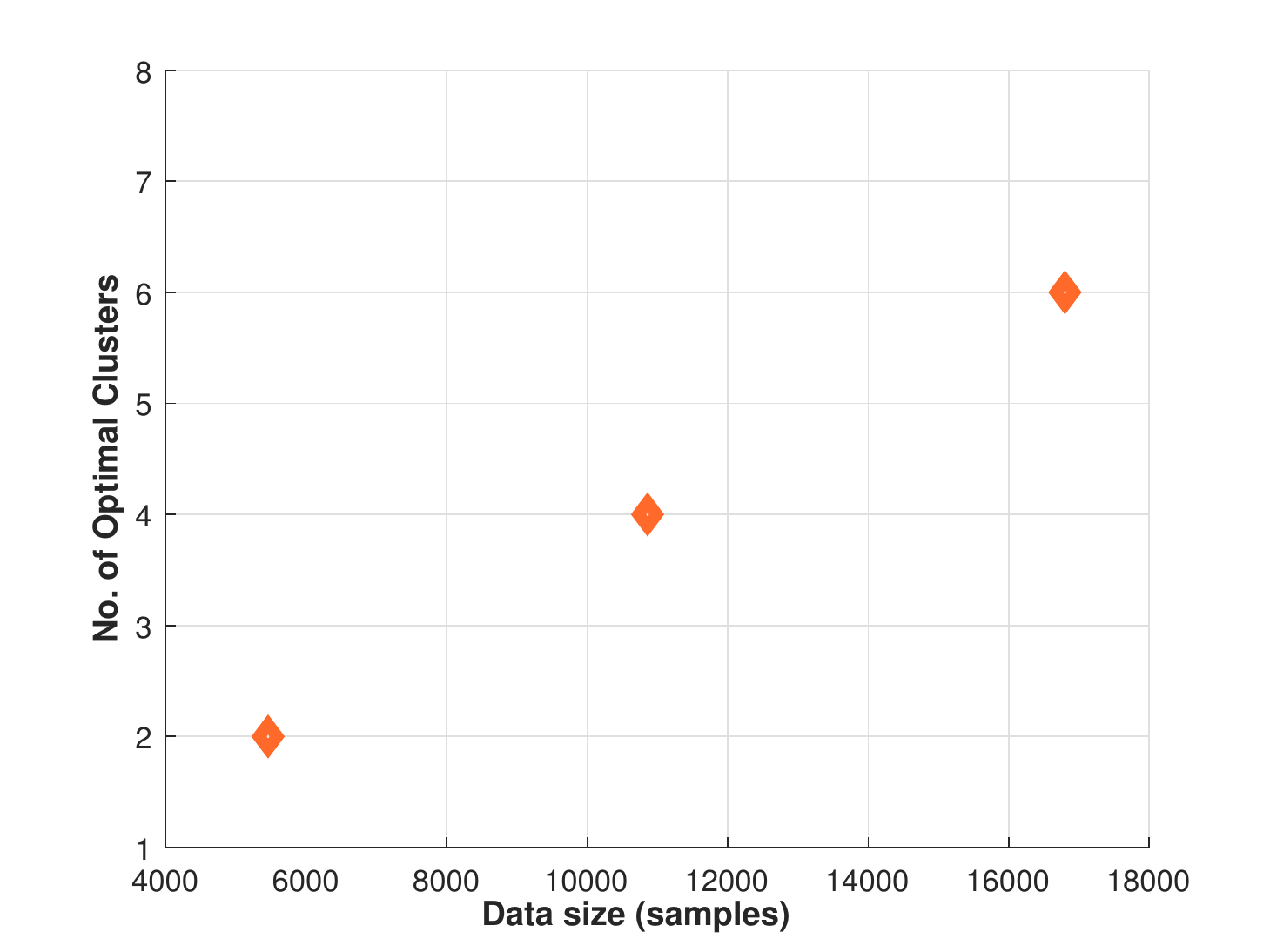}
\caption{Number of optimal clusters versus data size.}
\label{fig_optimal_cluster}
\end{figure}

One more information can be extracted from Fig. \ref{fig_nmse_comp_all}. It can be seen that the lowest NMSE is achieved for having different clusters for the three different scenarios. While the least NMSE for 5460 samples is achieved for $2$ clusters, the least NMSE is obtained for having $6$ clusters when considering Dataset-3 with 16800 location spots. We plot the number of optimal clusters versus the data size in Fig. \ref{fig_optimal_cluster}. For considering 5460 samples representing Dataset-1, we found $2$ disjoint clusters would suffice for the area of interest. But as we increase the number of observations in Dataset-2 with more location points, the area of interest demands $4$ disjoint clusters to provide the optimal result. Further extending the region with 16800 samples, the proposed solution requires $6$ optimal clusters that provides the lowest average NMSE. These results suggest that the required number of optimal clusters may change according to the size of the dataset and the considered environment. Generally, dividing the data samples into more clusters is equivalent to adding more codewords (or simply beams) in the codebook design for the given location points. As such, we are required to estimate the optimal number of clusters to design an efficient codebook. Our results suggest that the adopted codebook for Dataset-1 should have two unique codewords, while Dataset-2 and Dataset-3 require $4$ and $6$ codewords, respectively. With these results for learning the covariance matrices, a matched environment-aware beamforming codebook can be designed accordingly by adopting any of the existing techniques in~\cite{love_06, raghavan_07}.
%%%%%%%%%%%%%
%%%%%%%%%%%%%%%%%%%%%%%%%%%%%%%%%%%%%%%%%%%%%%%%%
%%%%%%%%%%%%%
\begin{figure}[h]
\centering
\includegraphics[width = .8\linewidth]{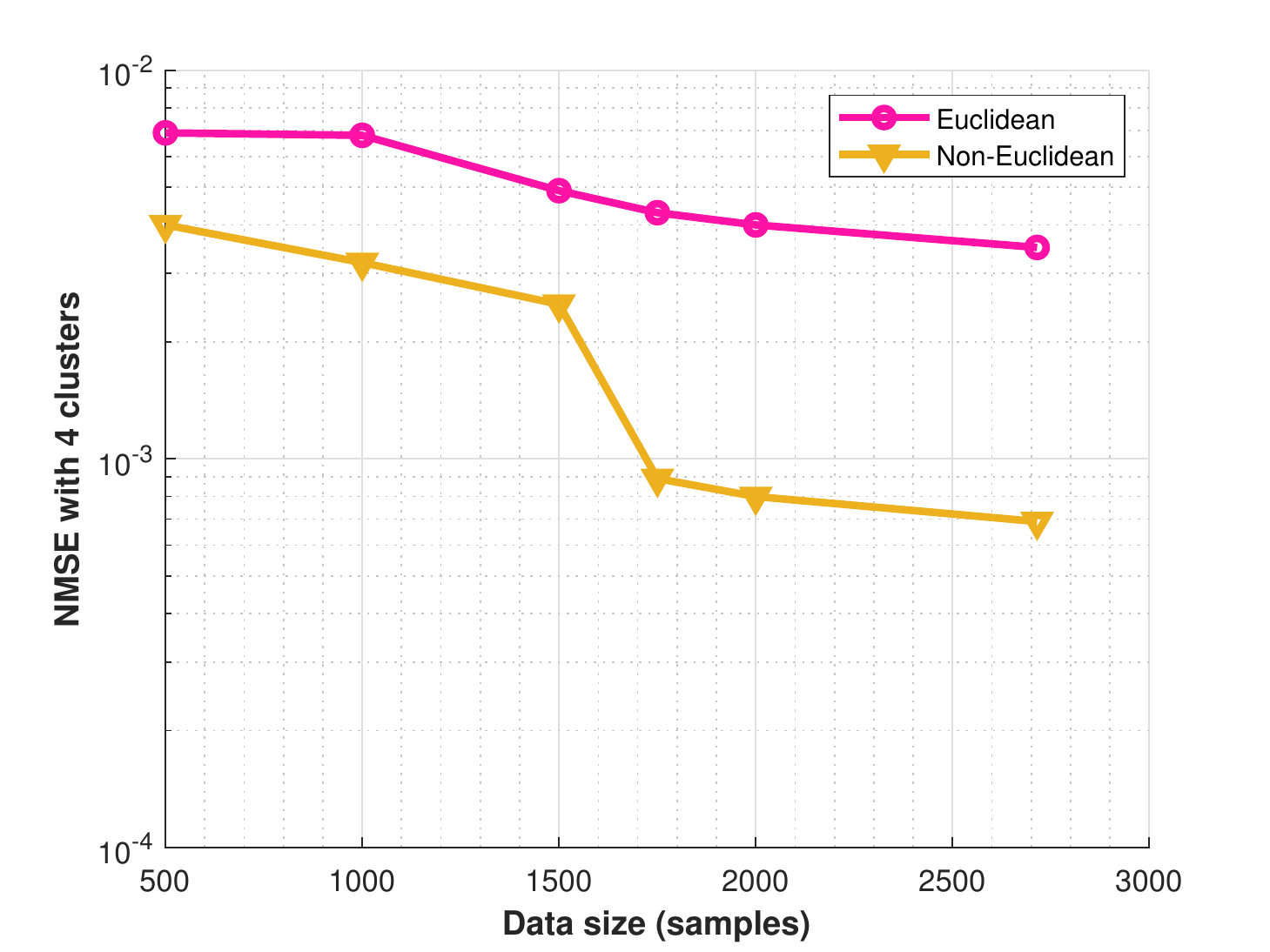}
\caption{Average NMSE for different regions of Dataset-2 having 4 clusters.}
\label{fig_fast}
\end{figure}

We are now interested to see how fast our proposed solution can cluster the given covariance matrices. To investigate this, we consider our adopted Dataset-2 and divided that into 4 disjoint groups. First, we consider 500 location points in each group (i.e., $4\times 500=2000$ samples from Dataset-2), and then gradually increased the group size to 2715 location points which is equivalent to considering the entire Dataset-2 with 10860 samples. Fig. \ref{fig_fast} shows the average NMSE having 4 clusters for considering different group size from Dataset-2. Considering 2715 samples in each clustered group, the proposed solution provides an NMSE of $7\times 10^{-4}$, but it increases to $8\times 10^{-4}$ while considering 2000 samples in each group (i.e., 8000 location points). 
As a result, the NMSE of the proposed solution increases by almost $15\%$ for reducing the data size by a total of 2860 samples.
In other words, our solution can make the estimation around $26\%$ faster by sacrificing only $15\%$ in learning the channel covariance matrices. Thus, the proposed learning scheme balances a trade-off between fast estimation and higher accuracy. 

%%%%%%%%%%%%%%%%%%%%%%%%%%%%%%%
\section{Conclusions}\label{sec_conclusion}
In this paper, we proposed a novel unsupervised G-ML scheme for learning the covariance matrices in a mmWave vehicular environment. The proposed solution does not require any prior labelled data for training and makes use of the underlying Riemannian-geometric characteristics of the spatially correlated  wireless channels while learning the covariance matrices. Simulation results suggest that the proposed non-euclidean based approach can provide up to $80\%$ higher accuracy compared to the existing euclidean-based schemes. Also, our proposed solution can efficiently determine the required number of codewords for a given vehicular scenario. In addition, the proposed G-ML strategy is capable to act $26\%$ faster with minimal sacrifice in the learning process.

%%%%%%%%%%%%%%%%%%%%%%%%%%%%%%%%%%%%%%%%%%%%%%%%%%%%%%%%%%%%%%%%%%%%%%%%%%%
% References
%%%%%%%%%%%%%%%%%%%%%%%%%%%%%%%%%%%%%%%%%%%%%%%%%%%%%%%%%%%%%%%%%%%%%%%%%%%
\balance

\end{document}